\documentclass[conference]{IEEEtran}
\IEEEoverridecommandlockouts
\usepackage{cite}
\usepackage{amsmath,amssymb,amsfonts}
\usepackage{algorithmic}
\usepackage{graphicx}
\usepackage{textcomp}
\usepackage{xcolor}
\usepackage{caption}
\usepackage{subcaption}
\usepackage{array}
\def\BibTeX{{\rm B\kern-.05em{\sc i\kern-.025em b}\kern-.08em
    T\kern-.1667em\lower.7ex\hbox{E}\kern-.125emX}}
\begin{document}

\title{%
		\vspace{-2em}
		\begin{center}
				\small This is an authors' copy of the paper to appear in Proceedings of the 35th IEEE International Workshop on Rapid System Prototyping (RSP 2024)
			\end{center}
		\vspace{0.3em}
		Cost-Effective Cyber-Physical System Prototype for Precision Agriculture with a Focus on Crop Growth
}

\author{\IEEEauthorblockN{Pawan Kumar}
\IEEEauthorblockA{
\textit{Arizona State University}\\
Tempe, AZ, United States \\
pkumar97@asu.edu }
\and
\IEEEauthorblockN{Hokeun Kim}
\IEEEauthorblockA{
\textit{Arizona State University}\\
Tempe, AZ, United States \\
hokeun@asu.edu}
}

\IEEEaftertitletext{\vspace{-1.45\baselineskip}}
\maketitle
\begin{abstract}
In precision agriculture, integrating advanced technologies is crucial for optimizing plant growth and health monitoring.
Cyber-physical system (CPS) platforms tailored to specific agricultural environments have emerged, but the diversity of these environments poses challenges in developing adaptive CPS platforms.
This paper explores rapid prototyping methods to address these challenges, focusing on non-destructive techniques for estimating plant growth.
We present a CPS prototype that combines sensors, microcontrollers, digital image processing, and predictive modeling to measure leaf area and biomass accumulation in hydroponic environments.
Our results show that the prototype effectively monitors and predicts plant growth, highlighting the potential of rapid CPS prototyping in promoting sustainability and improving crop yields at a moderate cost of hardware.
\end{abstract}

\section{Introduction}
The agricultural sector is transforming with the advent of advanced computing technologies that enable more precise and sustainable farming practices, namely, precision agriculture~\cite{pierce1999aspects}.
These innovations include non-destructive methods for estimating plant growth or disease detection as crucial tools for continuous and accurate plant health monitoring without causing damage~\cite{paturkar2020non, ang2022non}.
These methods are essential to maintaining the integrity of plants throughout their growth cycle, ensuring accurate assessments of growth patterns and overall health.
Digital image processing techniques have revolutionized the way plant growth measurements are conducted. 
Advanced sensors and automated systems can monitor these environmental conditions, providing real-time data that can be used to optimize growth conditions and improve plant health.

Integrating cyber-physical systems (CPS) into agriculture represents a significant advancement towards precision agriculture.
CPS enables real-time data collection, analysis, and decision-making, enhancing farm productivity and sustainability. By optimizing resource use and reducing labor costs, CPS contributes to the creation of intelligent farming environments that are both efficient and sustainable.
However, the current state-of-the-art CPS platforms for precision agriculture mostly target specific types of crops, and their usage is limited to particular environments and their requirements.
Therefore, computing platforms must be reengineered and customized for various environments, leading to delayed deployments and significant costs for development, upgrades, and maintenance.
For example, mass-production farming environments require large-scale controls using rovers or drones, with wireless, low-bandwidth, and intermittent communication. 

To address the challenges in diverse precision agriculture environments~\cite{junior2024precision, khosla2010precision}, a rapid prototyping method for CPS platforms is sorely needed.
This research aims to design and implement a preliminary cost-effective platform that leverages CPS to measure and predict environmental impacts on plant growth. By integrating various sensors, microcontrollers, and image processing techniques, we aim to develop a system that provides detailed morphological data critical for assessing plant growth.
Our goal is to explore prototyping methods that enhance the accuracy of growth measurements, promote more sustainable and efficient agricultural practices, and minimize the cost of prototype development.

\section{Related Work and Background}
\label{sec:background}
As the importance of platforms for precision agriculture rises, various CPS and Internet-of-Things (IoT) platforms and prototypes have been proposed.~\cite{vasisht2017farmbeats, popovic2017architecting, grimblatt2019precision, lanucara2020prototype, aleotti2018smart, benyezza2023smart, cimino2017low, fresco2018enhancing} over the recent years.
A variety of aspects of precision agriculture and CPS technologies have been leveraged.
FarmBeats~\cite{vasisht2017farmbeats} enables data collection on different types of sensors and hardware platforms using IoT technology.
Popovi{\'c} \textit{et al.}~\cite{popovic2017architecting} also uses the IoT for ecological monitoring; similarly, Grimblatt \textit{et al.}~\cite{grimblatt2019precision} build a prototype integrating the IoT for monitoring of small to medium size farms.
Lanucara  \textit{et al.}~\cite{lanucara2020prototype} propose a platform based on service-oriented architecture for precision agriculture.
Platforms and prototypes have been proposed targeting specific aspects of precision agriculture such as irrigation systems~\cite{aleotti2018smart} and greenhouse environments~\cite{benyezza2023smart}.
There have been prototype-based approaches focusing on low-cost and open-source availability~\cite{cimino2017low} and enhancing the effectiveness of precision agriculture~\cite{fresco2018enhancing}.


Non-destructive methods for estimating plant growth are crucial for continuous monitoring without harming the plant. Recent advancements focus on digital image processing to enhance the accuracy and efficiency of these methods. For instance, Islam \textit{et al.}~\cite{Islam2021}  and Zhang~\cite{Zhang2020} have developed algorithms to measure leaf area and dimensions in different plants using RGB and grayscale image transformations.
These methods provide vital data for analyzing plant health and growth patterns in a variety of settings, from agricultural fields to controlled environments like plant factories.

Technological advancements have extended into the monitoring and management of crop health, emphasizing the integration of sensors and data analytics in agriculture.
Kagalingan \textit{et al.}~\cite{Kagalingan2022} discuss the use of sensors and automated systems for controlling environmental conditions in indoor vertical gardens, demonstrating the potential of technology to replicate optimal growth conditions.
Additionally, Weraduwage \textit{et al.}~\cite{Weraduwage2015} present a model exploring the relationship between leaf area growth and biomass accumulation in \textit{Arabidopsis thaliana}\footnote{A small plant from the mustard family, also known as thale cress.}, showing how variations in carbon allocation can influence plant health.
Niklas~\cite{Niklas1994} contributes to this area by discussing the importance of modeling plant growth dynamics and biological scaling, providing a foundational understanding of the relationship between plant form and function.


The integration of CPS in farming and agriculture marks a significant shift towards more precise and automated farming practices. Recent projects like AFarCloud~\cite{Castillejo2020} and specialized models for intelligent husbandry farms~\cite{Chivarov2023} utilize CPS to enhance real-time data collection, analysis, and decision-making. These systems not only improve farm productivity and animal health but also contribute to sustainability by optimizing resource use and reducing labor costs. The developments in CPS highlight the future direction of agriculture, aiming for high efficiency and minimal environmental impact.

\section{Prototype Design and Implementation}
\label{sec:prototype}
In this section, we outline the design and implementation of our prototype, allowing rapid prototyping of a CPS platform for precision agriculture with a focus on crop growth.

\subsection{Prototype Design}

\begin{figure}
	\includegraphics[width=0.85\columnwidth]{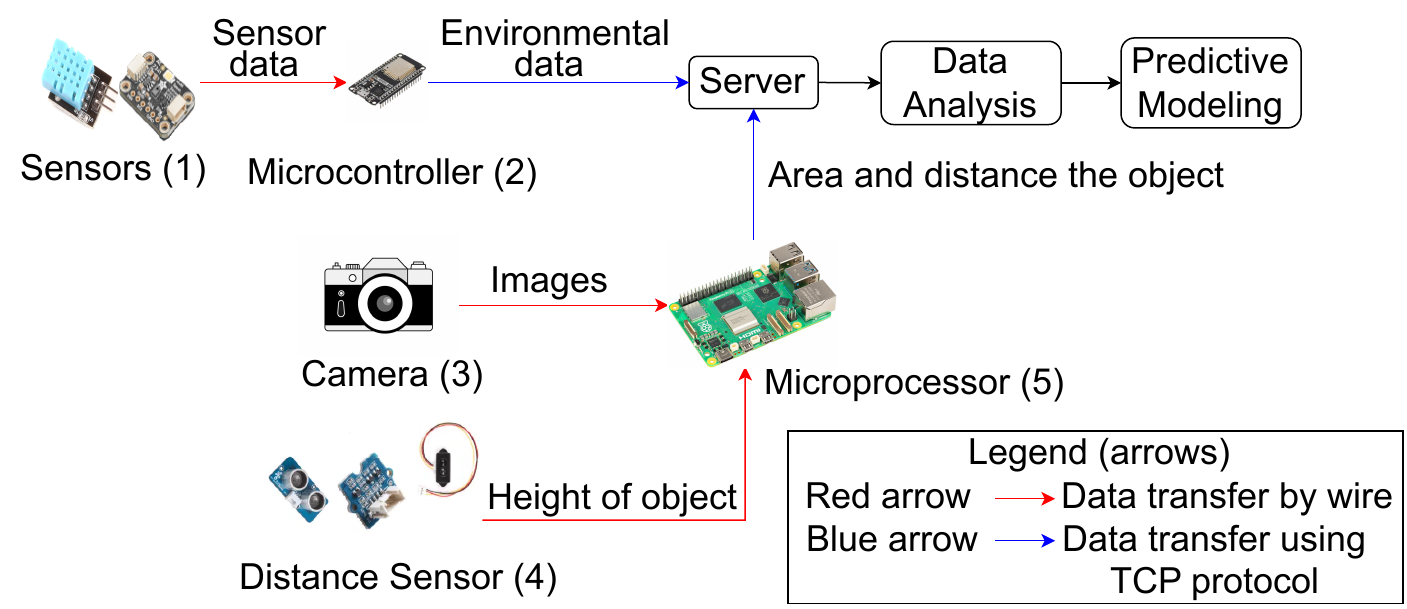}
	\caption{An overview of our prototype.}
	\label{fig:approach_diagram}
\end{figure}

\figurename~\ref{fig:approach_diagram} outlines an overview of our prototype for environmental monitoring and analysis, incorporating various technologies to measure and predict environmental impacts on objects within a specific area. At the cyber-physical interface of this prototype, distance sensor (shown in \figurename~\ref{fig:approach_diagram}--(4)) and a camera (shown in \figurename~\ref{fig:approach_diagram}--(3)) work together to collect images and sensors (shown in \figurename~\ref{fig:approach_diagram}--(1)) to collect environmental data.
These sensors (shown in \figurename~\ref{fig:approach_diagram}--(1)) can measure a wide range of environmental parameters, such as temperature, humidity, and light, providing a comprehensive dataset on the surrounding environment.

The collected sensor data and images are then processed by a microcontroller (shown in \figurename~\ref{fig:approach_diagram}--(2)) and a microprocessor (shown in \figurename~\ref{fig:approach_diagram}--(5)), respectively, for data handling and processing
The microcontroller performs initial data collection and preliminary processing, acting as a bridge to the more powerful computational resources in the server (or workstation), which further analyzes the images.

This processed information is subsequently sent to a remote server via TCP, indicating the use of remote computing or remote storage to aggregate and store the data.
The server plays a crucial role in the data analysis phase, where algorithms and models can be applied to interpret the collected data.
This analysis involves identifying the area and biomass of objects within the captured images, as well as their heights, possibly using machine learning (ML) techniques or other forms of predictive modeling to estimate future conditions or changes.

\subsection{Hardware Components}
\begin{figure}
	\centering
	\includegraphics[width=0.85\columnwidth]{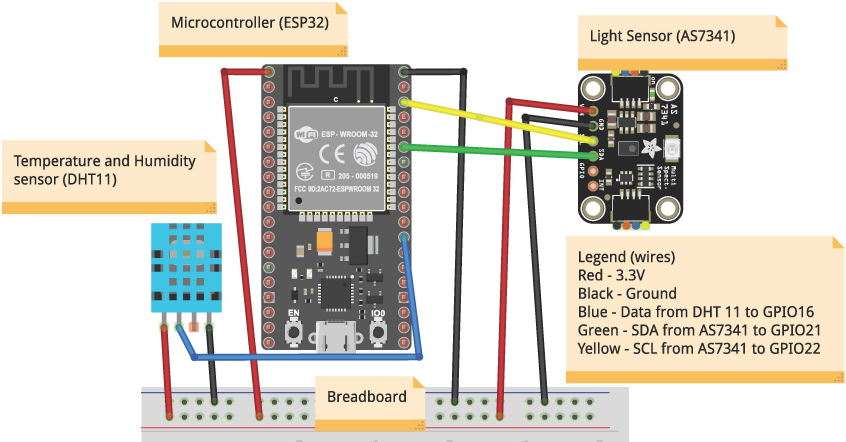}
	\caption{Hardware components diagram illustrating the setup of our prototype using ESP32, DHT11 and AS7341 with a legend represents the wire connections.}
	\label{fig:Circuit ESP32}
\end{figure}

 \figurename~\ref{fig:Circuit ESP32} shows the hardware components diagram that presents our prototype hardware setup combining an ESP32 \cite{CSESP32} microcontroller with two sensors: the DHT11 \cite{CSDHT11} for measuring temperature and humidity, and the AS7341 \cite{CSAS7341} for light sensing. The DHT11 sensor is connected to the ESP32 through a single data line to GPIO16, which is a standard approach for digital sensors that communicate through a single-wire protocol. Power to the DHT11 is supplied through a 3.3V connection, as indicated by the red wire, and a black wire serves as the common ground, which is essential for completing the circuit and allowing current to flow.

In contrast, the AS7341 sensor employs an I\(^2\)C communication protocol, necessitating two connections: the SDA (serial data line) and SCL (serial clock line), connected to GPIO21 and GPIO22 on the ESP32, respectively. The SDA line, marked with a green wire, allows for bidirectional data transfer between the sensor and the microcontroller, while the yellow SCL wire provides the clock signal that synchronizes data transmission. This configuration is typical for I\(^2\)C devices, where multiple sensors can share the same SDA and SCL lines if needed, allowing for an expandable sensor array. Power is again supplied by a 3.3V connection, and a shared ground completes the necessary power circuit for the AS7341.

\subsection{Hardware Prototype Design}
\begin{figure}
	\centering
	\includegraphics[width=0.75\columnwidth]{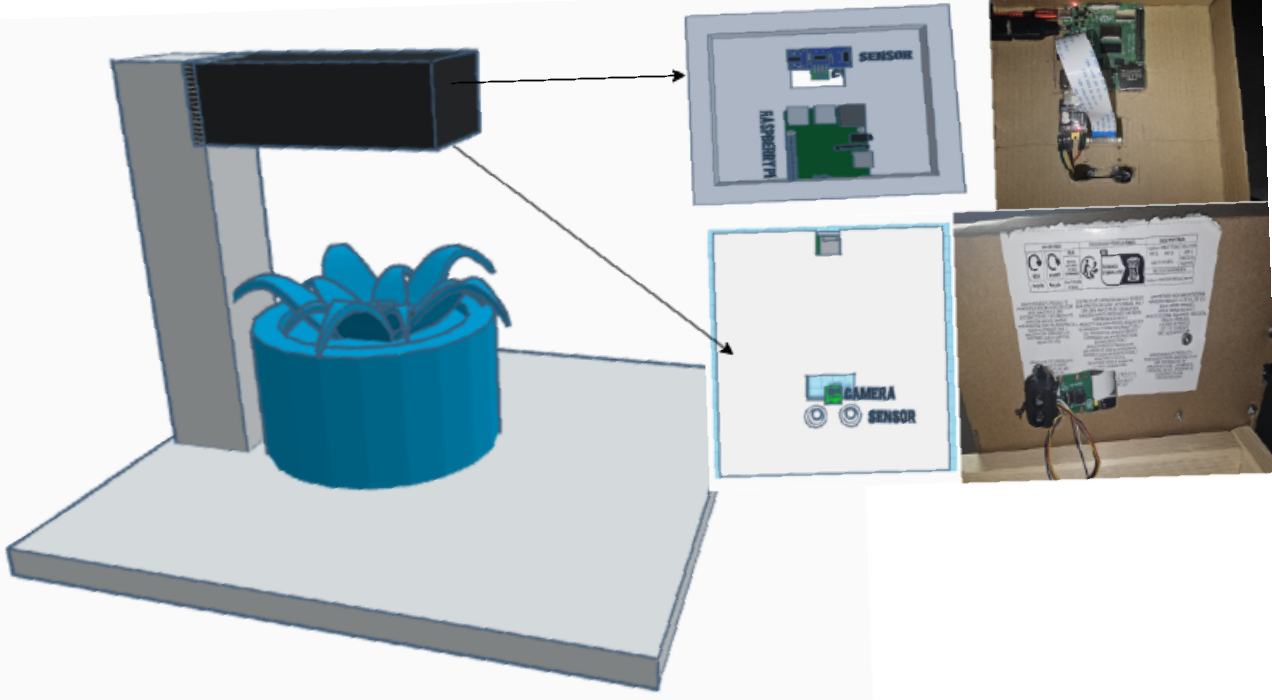}
	\caption{Support structure to hold the hardware components above the plant.}
	\label{fig:support_structure}
\end{figure}
 \figurename~\ref{fig:support_structure} shows our prototype's support structure, which we engineered to maintain the data acquisition hardware at a height of 13 inches from the base.
The gray skeletal part provides stable support for the black hardware setup (shown in \figurename~\ref{fig:support_structure}) , ensuring the precise positioning of the sensors and camera.
 
\begin{figure}
	\centering
	\includegraphics[width=0.5\textwidth]{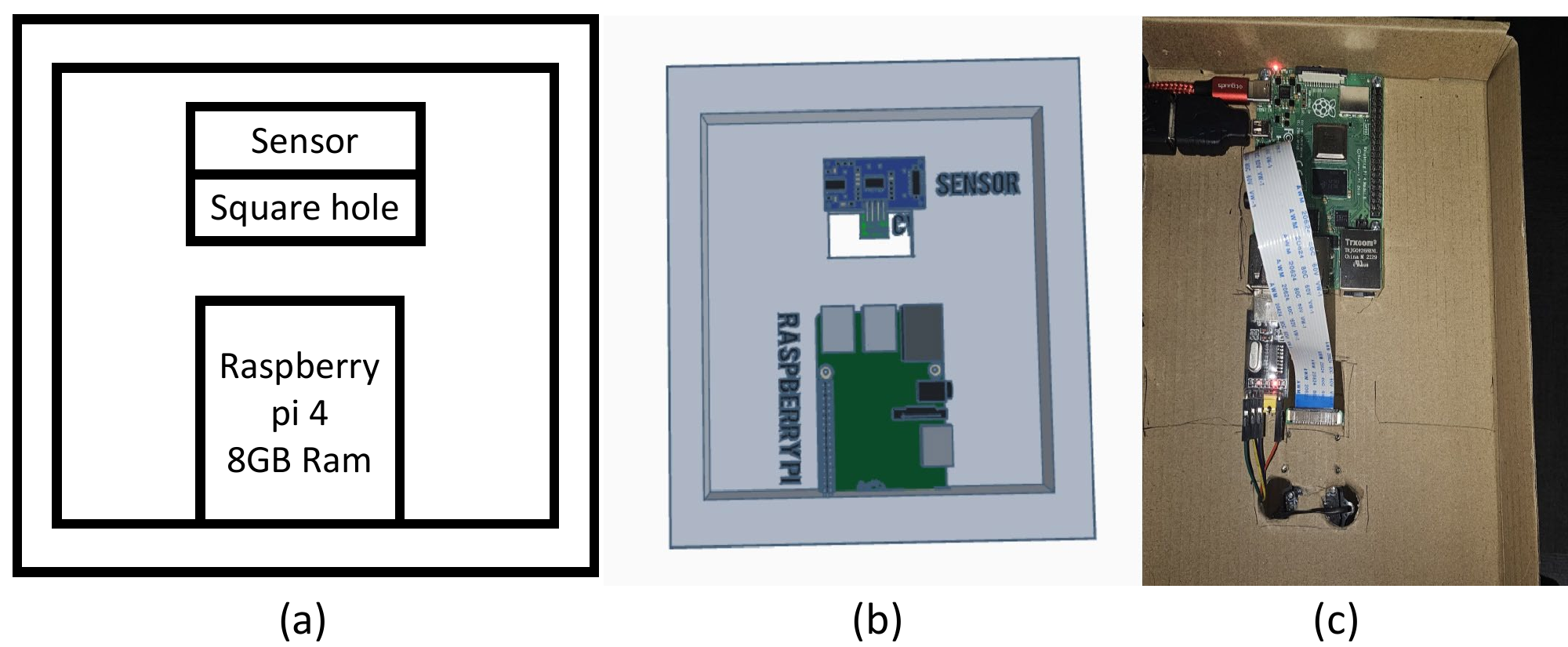}
	\caption{Internal hardware configuration consisting of Raspberry Pi~\cite{CSRpi}, camera~\cite{CSRpiCam}, and distance sensor, presented as (a) a 2D image, (b) a 3D image, and (c) an actual picture.}
	\label{fig:Internal Hardware Configuration}
\end{figure}

The internal hardware setup of our prototype shown in  \figurename~\ref{fig:Internal Hardware Configuration} includes a microprocessor that is interfaced with a distance sensor for measuring plant height, a key variable for biomass estimation.
This setup captures detailed morphological data critical for assessing plant growth.

\begin{figure}
	\centering
	\includegraphics[width=0.5\textwidth]{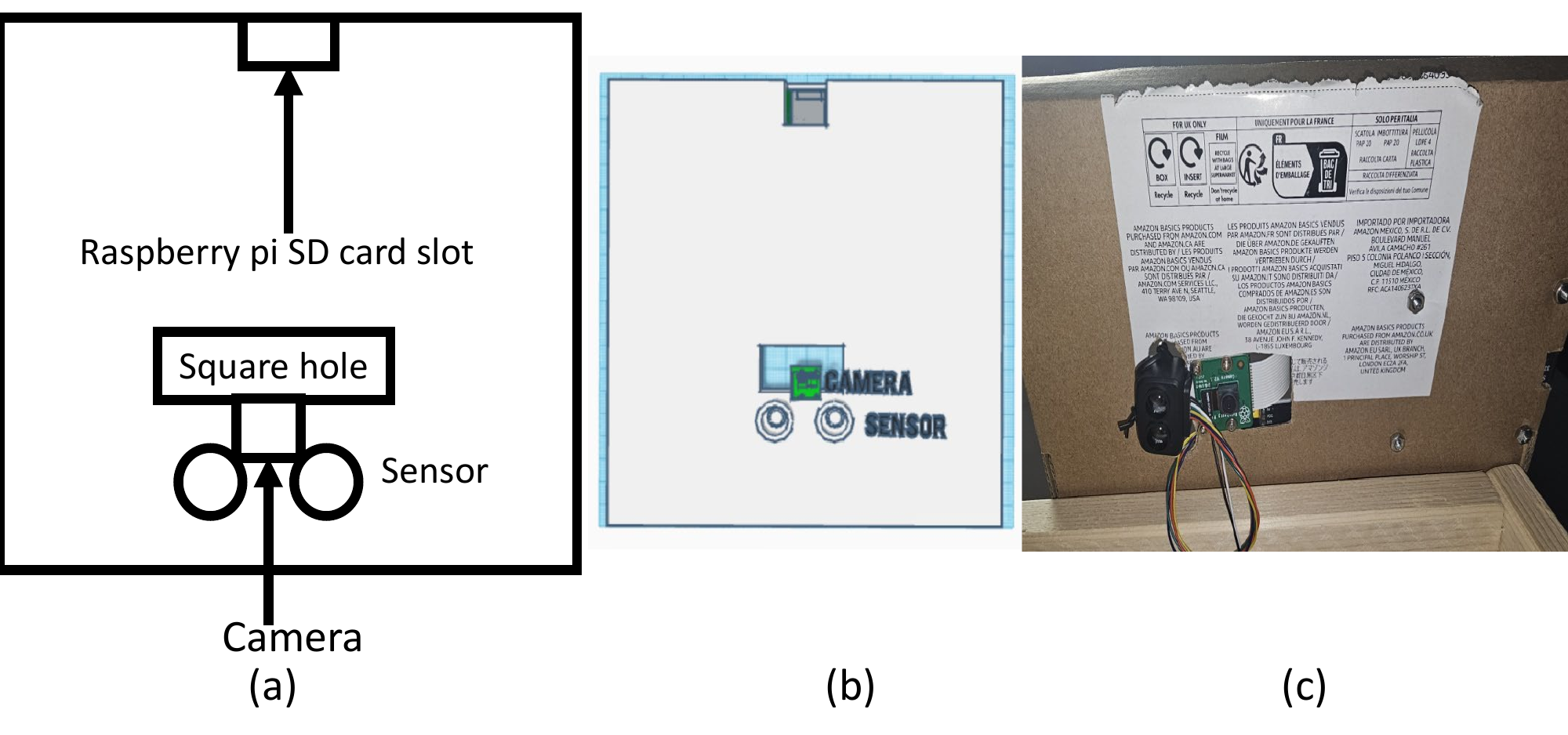}
	\caption{Distance sensor and camera placed close to each other to measure the distance and take images respectively presented as (a) a 2D image, (b) a 3D image, and (c) an actual picture.}
	\label{fig:Camera and Sensor Placement}
\end{figure}

In  \figurename~\ref{fig:Camera and Sensor Placement}, we attached the camera strategically close to the microprocessor, alongside the distance sensor components. This arrangement facilitates the simultaneous capture of visual and spatial data, allowing for a comprehensive analysis of the plant's physical characteristics.

\section{Crop Growth Estimation}
\label{Area Measurement}

\begin{table}
	 \begin{tabular}{|l|>{\raggedleft\arraybackslash}p{0.6cm}|>{\raggedleft\arraybackslash}p{1.45cm}|>{\raggedleft\arraybackslash}p{1.12cm}|>{\raggedleft\arraybackslash}p{1.05cm}|>{\raggedleft\arraybackslash}p{1cm}|}
		\hline
		Object & Actual area (cm\(^2\)) & Area measured by ultrasonic sensor (cm\(^2\)) & Error of ultrasonic sensor (\%)  & Area measured by TF-Luna (cm\(^2\)) &\% error of TF-Luna \\
		\hline
		2$\times$2 cube & 25.00 & 25.63 & 2.49 & 24.50 & 2.02 \\
		Leaf 1 & 3.42 & 2.22 & 53.89 & 3.04 & 12.19 \\
		Leaf 2 & 15.78 & 10.34 & 52.65 & 14.71 & 7.27 \\
		Leaf 3 & 18.98 & 14.88 & 27.55 & 20.16 & 5.84 \\
		Leaf 4 & 7.85 & 4.46 & 76.09 & 9.53 & 17.63 \\
		\hline
	\end{tabular}
	\caption{Measurement results using an ultrasonic sensor (HC-SR04)~\cite{CSHCSR04} and a TF-Luna sensor~\cite{CSTF-LUNA} with actual area measured using vernier caliper.}
	\label{table:actual_area}
\end{table}

We measured and estimated crop growth using the leaf area of plants as discussed in Section~\ref{sec:background}.
The evaluation of area measurements using different sensors is presented in \tablename~\ref{table:actual_area}, which combines the results from both the ultrasonic sensor (HC-SR04)~\cite{CSHCSR04} and the TF-Luna sensor\footnote{TF-Luna is a low-cost,single-point LiDAR sensor utilizing the time-of-flight technology for accurate distance measurement (0.2-8m) with 1cm resolution and $\pm$6cm accuracy.}. This table provides a detailed assessment of area measurements for a 2$\times$2 cube of dimensions 5cm $\times$ 5cm and four different leaves from different plants, specifically \textit{Portulacaria afra}\footnote{A succulent plant with a reddish stem and small green leaves.} (Leaf 1), \textit{Scutellaria baicalensis}\footnote{A flowering plant, also known by its common name, Baikal skullcap.} (Leaf 2), \textit{Capsicum annuum `jalapeño'}\footnote{a fruiting plant, also known as sweet and chili pepper.} (Leaf 3), and \textit{Ocimum basilicum}\footnote{A culinary herb, also called basil or great basil.} (Leaf 4), comparing the performance of the two sensors against actual areas measured using a vernier caliper.

For the 2$\times$2 cube, the ultrasonic sensor measured an average area of 25.63cm\(^2\), deviating from the actual area of 25.00cm\(^2\) by a percentage error of 2.46\%. The TF-Luna sensor measured an average area of 24.50cm\(^2\), resulting in a lower percentage error of 2.02\%. This showcases the high precision of the TF-Luna sensor in measuring simple geometries.

For the leaves, the ultrasonic sensor showed significant discrepancies. Leaf 1’s average measured area was 2.22cm\(^2\) against an actual area of 3.42cm\(^2\), resulting in a high percentage error of 53.89\%. Leaf 2’s average measured area of 10.34cm\(^2\) with a percentage error of 52.65\% compared to the actual area of 15.78cm\(^2\). Leaf 3’s measured area averaged 14.88cm\(^2\), with a percentage error of 27.55\% compared to the actual area of 18.98cm\(^2\). Leaf 4 demonstrated the greatest error, with an average measured area of 4.46cm\(^2\) against an actual area of 7.85cm\(^2\), with a percentage error of 76.09\%.

In contrast, the TF-Luna sensor provided more accurate and consistent measurements. Leaf 1’s average measured area was 3.05cm\(^2\) against an actual area of 3.42cm\(^2\), resulting in a lower percentage error of 12.19\%. Leaf 2’s measurements indicated an average area of 14.71cm\(^2\) with a percentage error of 7.27\% compared to the actual area of 15.78cm\(^2\). Leaf 3’s measured area averaged 20.16cm\(^2\), with a percentage error of 5.84\% compared to the true area of 18.98cm\(^2\). Leaf 4 showed a measured area of 9.53cm\(^2\) against an actual area of 7.85cm\(^2\), resulting in a percentage error of 17.63\%.

The data from \tablename~\ref{table:actual_area} clearly indicates that the TF-Luna sensor has substantially lower percentage errors across various samples, suggesting a higher degree of precision and reliability. The TF-Luna sensor proves to be more suitable for accurately measuring areas of both simple and complex geometries, as evidenced by the lower percentage errors compared to the ultrasonic sensor. The TF-Luna’s enhanced measurement consistency and accuracy make it a preferable option for tasks requiring high precision and fine detail. Consequently, the TF-Luna sensor was selected for use in further measurements and analyses.

\begin{figure}
	\centering
	\begin{subfigure}{0.32\columnwidth}
		\includegraphics[width=\columnwidth]{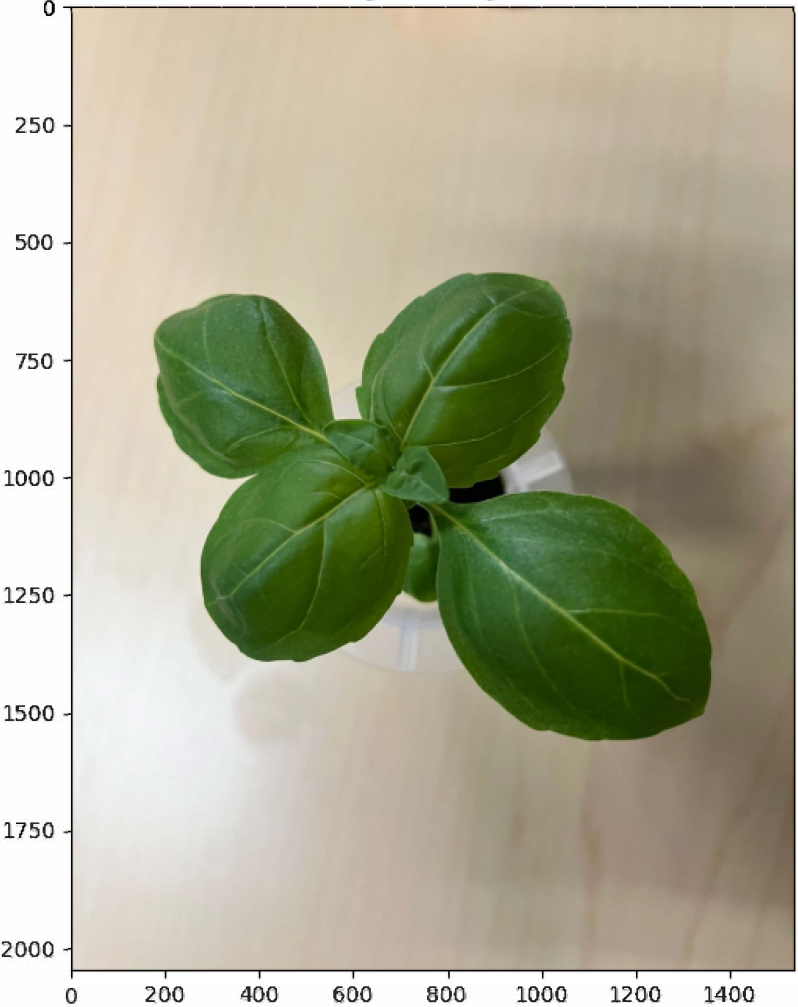}
		\caption{An original image of a basil plant taken by camera.}
		\label{fig:org_img}
	\end{subfigure}
	\hfill
	\begin{subfigure}{0.32\columnwidth}
		\centering
		\includegraphics[width=\columnwidth]{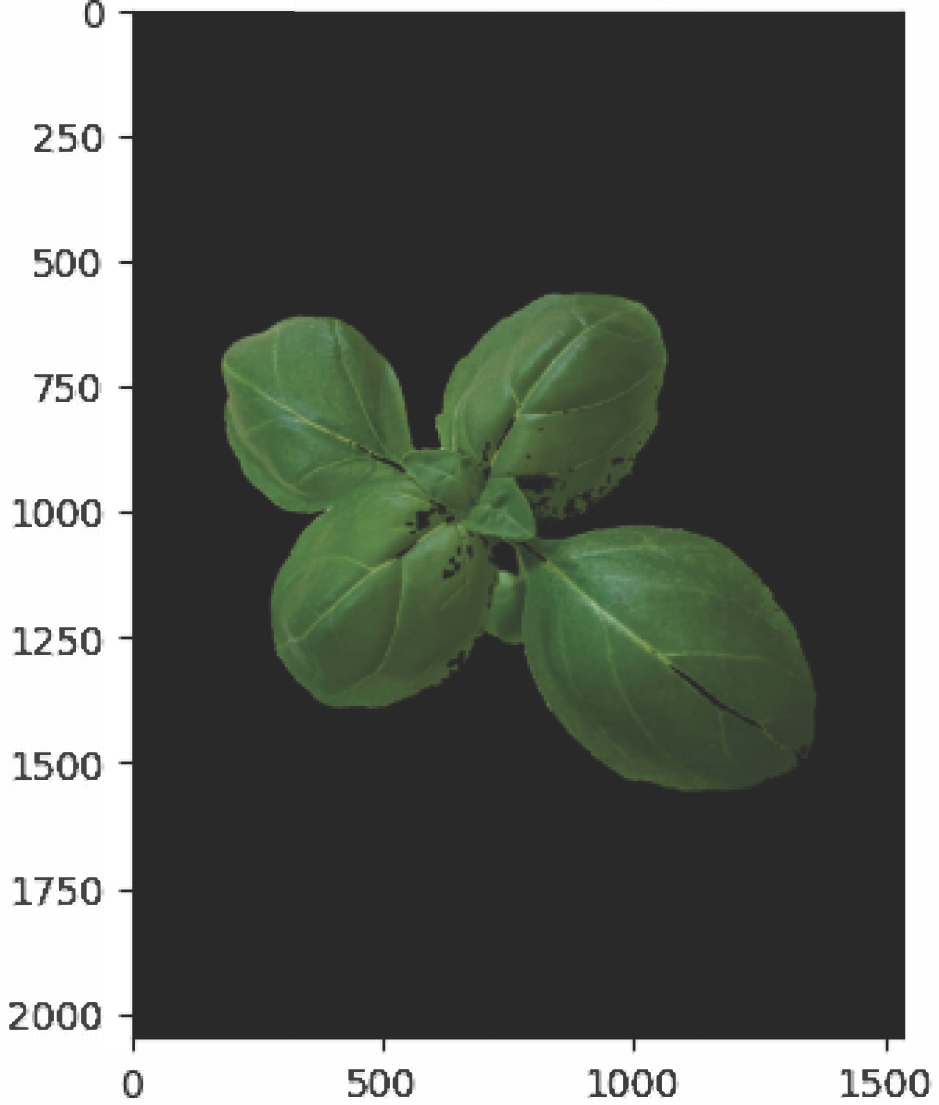}
		\caption{A processed image after green color thresholding.}
		\label{fig:green_image}
	\end{subfigure}
	\hfill
	\begin{subfigure}{0.32\columnwidth}
		\centering
		\includegraphics[width=\columnwidth]{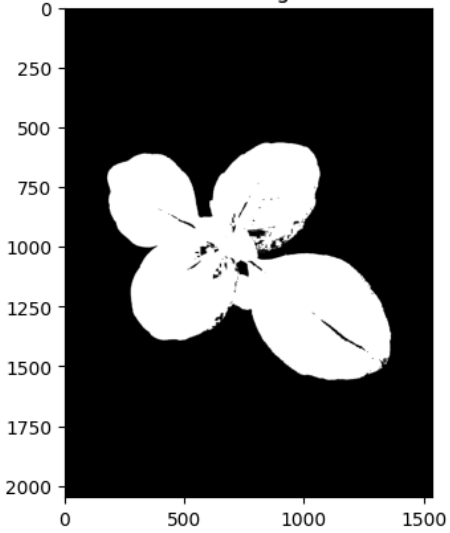}
		\caption{A final image after masking the image in  \figurename~\ref{fig:green_image}.}
		\label{fig:mask_image}
	\end{subfigure}
	\caption{Measuring the number of pixels from the top view of a basil plant (binomial name: \textit{Ocimum basilicum}).}
	\label{fig:pixel_image}
\end{figure}

To measure the leaf area of the plant, we follow a structured image processing workflow, enhanced by using reference objects and distance adjustments. Initially, an image of the plant is captured by the camera, as shown in \figurename~\ref{fig:org_img}. This image is processed to enhance its quality and clarity, ensuring that the plant's features are well-defined. The next step involves isolating the target area, specifically the green part of the leaves, using color thresholding techniques. \figurename~\ref{fig:green_image} shows the result of this step, where only the green regions, which represent the leaves, are highlighted, ensuring that non-leaf elements are excluded from the analysis.

Following this, a mask is applied to the image to clearly differentiate the leaf area from the background. \figurename~\ref{fig:mask_image} demonstrates this process, where the leaf area is isolated by applying a binary mask that effectively separates the leaf pixels from the non-leaf pixels. In \figurename~\ref{fig:mask_image}, the masked image is analyzed to determine the leaf area in pixels by counting the number of pixels that fall within the masked region. For instance, the green area of the image is calculated as 680,505.0 pixels.

	\begin{figure}
		\centering
		\begin{subfigure}{0.48\columnwidth}
			\centering
			\includegraphics[width=0.75\columnwidth]{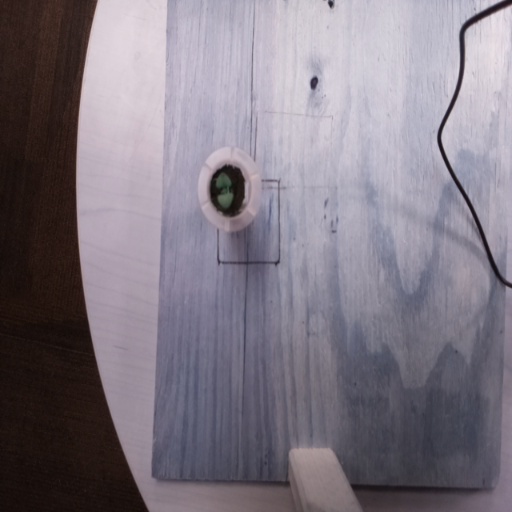}
			\caption{A plant far from the camera.}
			\label{fig:far_image}
		\end{subfigure}
		\hfill
		\begin{subfigure}{0.48\columnwidth}
			\centering
			\includegraphics[width=0.75\columnwidth]{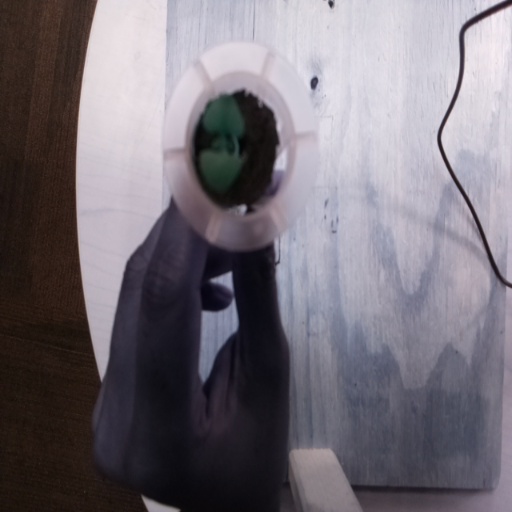}
			\caption{A plant near the camera.}
			\label{fig:near_image}
		\end{subfigure}
		\caption{Plant images at different distances from the camera.}
		\label{fig:cam_fig}
	\end{figure}

To accurately measure the leaf area, we used a reference object with a known area. The base ratio is calculated using the formula:
\begin{equation}
	\text{base\_ratio} = \frac{\text{reference\_object\_pixel\_area}}{\text{reference\_object\_real\_area\_cm}^2}
	\label{eq:base_ratio}
\end{equation}

  This ratio helps in converting pixel measurements to real-world units. The area is then computed to account for any changes in the distance between the camera and the object, using the formula:
\begin{equation}
	\text{area} = \text{base\_ratio} \times (\frac{\text{current\_distance}}{\text{reference\_distance}})^2 
	\label{eq:distance_adjusted_ratio}
\end{equation}

Since the width and height of the object scale linearly with distance, the distance ratio is squared to adjust the measurements accurately. In this example, the area of the green object is calculated as 259.93cm\(^2\) by converting the pixel count of 680,505 pixels using the distance-adjusted ratio.

In \figurename~\ref{fig:far_image}, the plant is placed at a distance from the camera, while in \figurename~\ref{fig:near_image}, the plant is closer to the camera. The formula is implemented to ensure that the images taken by the camera in \figurename~\ref{fig:far_image} and \figurename~\ref{fig:near_image} yield the same leaf area measurement, regardless of the distance. This comprehensive approach ensures that accurate measurements are obtained regardless of the camera's position relative to the plant, as outlined in the accompanying algorithm. The resulting data, representing the leaf area in square centimeters, is recorded and utilized for further analysis.

\tablename~\ref{table:tfluna} and \tablename~\ref{table:ultrasonic} present the results of measuring leaf area using two different sensors: the TF-Luna sensor and the ultrasonic sensor.
We took measurements to evaluate the accuracy and effectiveness of these sensors when used with the image processing method described earlier, involving steps from \figurename~\ref{fig:pixel_image} for capturing, processing, and analyzing the images. We used basil (\textit{Ocimum basilicum}) for the rest of the experiments.

\tablename~\ref{table:tfluna} summarizes the data collected using the TF Luna sensor. The ``Actual distance (AD)'' is consistently 22.00cm across all measurements. The ``Measured distance (MD)'' varies slightly from 24.00 to 25.00cm, resulting in a percentage error in the distance measurement ranging from 8.33\% to 12\%. The leaf area calculated using the AD and the MD is provided, showing small differences between the two. The percentage error in the area measurement using the TF Luna sensor is relatively low, with an average error of 6.82\%, indicating that this sensor provides reasonably accurate distance and area measurements.

\tablename~\ref{table:ultrasonic} shows the data collected using the ultrasonic sensor. Similar to Table IV, the AD is 22.00cm. However, the MD exhibits larger variations, ranging from 29.03\% to 42.11\%, leading to a significantly higher percentage error in the distance measurement, averaging 34.91\%.
This larger error impacts the leaf area calculation. The area calculated using the AD and the MD shows substantial differences, with the percentage error in the area measurement using the ultrasonic sensor averaging 82.91\%. This high error rate suggests that the ultrasonic sensor is less accurate in measuring distances and, consequently, in determining the leaf area.

In conclusion, the TF Luna sensor demonstrates a lower average percentage error in both distance and area measurements compared to the ultrasonic sensor. The TF Luna sensor's average percentage error in area measurement is significantly lower at 6.82\%, compared to the ultrasonic sensor's 82.91\%, highlighting its superior performance in this application. These results suggest that the TF Luna sensor is more suitable for accurate leaf area measurements using the described image processing method.

\begin{table}
	\centering
	\begin{tabular}{|>{\raggedleft\arraybackslash}p{1cm}|>{\raggedleft\arraybackslash}p{1cm}|>{\raggedleft\arraybackslash}p{1cm}|>{\raggedleft\arraybackslash}p{1cm}|>{\raggedleft\arraybackslash}p{1cm}|>{\raggedleft\arraybackslash}p{1cm}|}
		\hline
		Actual distance (AD) in cm & Measured distance (MD) in cm &\% error in difference & Area using AD in cm\(^2\) & Area using MD in cm\(^2\) &\% error in area \\
		\hline
		22.00 & 24.00 & 8.33 & 1.28 & 1.46 & 12.33 \\
		22.00 & 25.00 & 12.00 & 1.35 & 1.42 & 4.93 \\
		22.00 & 25.00 & 12.00 & 1.24 & 1.31 & 5.34 \\
		22.00 & 25.00 & 12.00 & 1.24 & 1.33 & 6.77 \\
		22.00 & 25.00 & 12.00 & 1.3 & 1.22 & 6.56 \\
		22.00 & 24.00 & 8.33 & 1.50 & 1.66 & 9.64 \\
		22.00 & 25.00 & 12.00 & 1.35 & 1.53 & 11.76 \\
		22.00 & 24.00 & 8.33 & 1.44 & 1.42 & 1.41 \\
		22.00 & 24.00 & 8.33 & 1.33 & 1.40 & 5.00 \\
		22.00 & 24.00 & 8.33 & 1.51 & 1.58 & 4.43 \\
		\hline
		\multicolumn{2}{|c|}{Average \% error} & 10.17 & \multicolumn{2}{|c|}{Average \% error} & 6.82 \\
		\hline
	\end{tabular}
	\caption{Measurement results using TF Luna sensor.}
	\label{table:tfluna}
\end{table}

\begin{table}
	\centering
	\begin{tabular}{|>{\raggedleft\arraybackslash}p{1cm}|>{\raggedleft\arraybackslash}p{1cm}|>{\raggedleft\arraybackslash}p{1cm}|>{\raggedleft\arraybackslash}p{1cm}|>{\raggedleft\arraybackslash}p{1cm}|>{\raggedleft\arraybackslash}p{1cm}|}
		\hline
		Actual distance (AD) in cm & Measured distance (MD) in cm &\% error in difference & Area using AD in cm\(^2\) & Area using MD in cm\(^2\) &\% error in area \\
		\hline
		22.00 & 35.00 & 37.14 & 1.56 & 8.04 & 80.60 \\
		22.00 & 32.00 & 31.25 & 1.38 & 7.00 & 80.29 \\
		22.00 & 33.00 & 33.33 & 1.56 & 7.73 & 79.82 \\
		22.00 & 32.00 & 31.25 & 1.25 & 7.45 & 83.22 \\
		22.00 & 34.00 & 35.29 & 1.44 & 7.33 & 80.35 \\
		22.00 & 35.00 & 37.14 & 1.33 & 8.02 & 83.42 \\
		22.00 & 34.00 & 35.29 & 1.35 & 7.67 & 82.4 \\
		22.00 & 38.00 & 42.11 & 1.52 & 13.35 & 88.61 \\
		22.00 & 36.00 & 38.89 & 1.46 & 12.56 & 88.38 \\
		22.00 & 33.00 & 33.33 & 1.40 & 7.67 & 81.75 \\
		22.00 & 31.00 & 29.03 & 1.35 & 8.03 & 83.19 \\
		\hline
		\multicolumn{2}{|c|}{Average \% error} & 34.91 & \multicolumn{2}{|c|}{Average \% error} & 82.91\\
		\hline
	\end{tabular}
	\caption{Measurement results using ultrasonic sensor.}
	\label{table:ultrasonic}
\end{table}


\section{Crop Growth Prediction Model}
According to Niklas~\cite{Niklas1994}, the use of log transformation can simplify complex relationships between variables in plants. This method helps convert non-linear relationships into linear forms, facilitating easier analysis and interpretation.

The key formulas used in our analysis are:

\textbf{1. Power Function Equation:}
\begin{equation}
	Y_1 = \beta Y_2^{\alpha}
	\label{eq:Power Function}
\end{equation}

In the equation~(\ref{eq:Power Function}), $Y_1$
represents the dependent variable (e.g., biomass accumulation), $Y_2$
represents the independent variable (e.g., leaf area), $\alpha$ is the scaling exponent that indicates how changes in $Y_2$ affect $Y_1$, and 
$\beta$ is a scaling coefficient. 
This equation suggests a non-linear relationship between the biomass accumulation and leaf area.

\textbf{2. Log-Transformed Linear Equation:}
\begin{equation}
	\log Y_1 = \log \beta + \alpha \log Y_2
	\label{eq:log_formula}
\end{equation}

By taking the logarithm of both sides of the power function equation, we obtain a linear relationship (equation (\ref{eq:log_formula})) where $\log Y_1$ is the dependent variable, $\log Y_2$ is the independent variable, $\alpha$ is the slope of the line, and $\log \beta $ is the intercept. This transformation simplifies the complexity of the non-linear relationship, making it easier to analyze and interpret using linear regression techniques.

We chose to use linear regression and Bayesian linear regression for several reasons. Linear regression, particularly on log-transformed data, simplifies the complexity of non-linear relationships by converting them into linear forms. This transformation allows for easier assessment of statistical significance and more straightforward interpretation of the parameters. By log-transforming the data, we can linearize power functions, which facilitates a robust understanding of the underlying biological relationships.

Using linear regression on log-transformed data helps to account for variability and provides a more robust understanding of these relationships. It simplifies the modeling process and makes it easier to analyze the data. Furthermore, Bayesian linear regression offers a probabilistic framework that effectively handles uncertainties in the data, providing more robust and reliable predictions under varying conditions. This method allows us to incorporate prior knowledge and quantify the uncertainty in our estimates, enhancing the interpretability and predictive power of our models.

Thus, the combination of these two approaches allows for a comprehensive analysis, balancing simplicity and statistical rigor with probabilistic insights.

\section{Evaluation}
This chapter evaluates our research, 
mainly from the perspectives of cost, accuracy, and practicality of the prototype.

\subsection{Cost Evaluation}
The total cost, as detailed in \tablename~\ref{table:Table_Cost}, amounts to a modest \$202.35.
This affordability of the prototype will allow research without a significant financial burden.
A crucial decision in the design was the choice of the TF-Luna sensor over a traditional ultrasonic sensor. While the TF-Luna carries a higher price point at \$25.98, its inclusion is justified by the superior accuracy, which is explained in Section~\ref{Area Measurement}. Therefore, while keeping the overall cost low, the system does not skimp on the reliability and precision of its components, illustrating a balanced approach to budgeting without sacrificing performance.
Note that the cost of wood and screws used in the construction of the prototype is not included in the total, as these materials were provided by Arizona State University.

\begin{table}
	\centering
	\begin{tabular}{|p{4cm}|>{\raggedleft\arraybackslash}p{2.4cm}|p{1.2cm}|}
		\hline
		Equipment & Cost per unit (USD) & References \\
		\hline
		GardenCube & 42.56 & \cite{CSGardencube} \\
		Raspberry Pi 4 8GB & 74.68 & \cite{CSRpi} \\
		ESP32 DEVBOARD-J & 8.99 & \cite{CSESP32} \\
		TF-Luna (LIDAR sensor) & 25.98 & \cite{CSTF-LUNA} \\
		AS7341 (Light sensor) & 15.95 & \cite{CSAS7341} \\
		DHT11 & 2.23 & \cite{CSDHT11} \\
		Raspberry Pi Camera V2 & 14.99 & \cite{CSRpiCam} \\
		Jumper wires & 6.98 &  \\
		USB-C cable & 9.99 & \\
		\hline
		Total & 202.35 & \\
		\hline
	\end{tabular}
	\caption{Cost evaluation of the hardware equipment used for constructing our CPS prototype.}
	\label{table:Table_Cost}
\end{table}

\subsection{Comparative Analysis of Modeling Approaches}

\begin{table}
	\centering
	\begin{tabular}{|l|>{\raggedleft\arraybackslash}p{2cm}|>{\raggedleft\arraybackslash}p{2cm}|}
		\hline
		Dataset type & Mean squared error (MSE) & R-squared value \\
		\hline
		Validation & 171.37 & 0.91 \\
		\hline
		Test & 156.12 & 0.93\\
		\hline
	\end{tabular}
	\caption{Linear regression model metrics.}
\label{table:Table_Linear}
\end{table}

\begin{table}
	\centering
	\begin{tabular}{|l|>{\raggedleft\arraybackslash}p{2cm}|>{\raggedleft\arraybackslash}p{2cm}|}
		\hline
		Dataset type & Mean squared error (MSE) & R-squared value \\
		\hline
		Validation & 185.61 & 0.90 \\
		\hline
		Test & 169.90 & 0.92\\
		\hline
	\end{tabular}
	\caption{Bayes Linear Regression Model Metrics.}
\label{table:Table_Bayes}
\end{table}

To effectively evaluate the prediction model, we utilized datasets comprising environmental factors, leaf area measurements, and plant weights. The environmental data was collected using DHT11 and AS7341 sensors, capturing temperature, humidity, and spectral data every hour. The dataset includes over 1,100 rows of hourly data points. Leaf area measurements and plant weights were recorded once a day.

The leaf area dataset includes measurements for five different basil plants (B1-B5), capturing the distance using our prototype explained in Section~\ref{sec:prototype} and the calculated leaf area in square centimeters.
The plant weight dataset includes both the measured and actual weights of the plants, with the actual weight of the plant derived by subtracting the known weight of the sponge and basket from the measured weight.
Data for plants B1 to B3 was collected from March 25, 2024, to April 3, 2024, and for plants B4 to B5 from April 24, 2024, to June 4, 2024.
For model training and evaluation, the datasets were preprocessed and split into training (60\%), validation (20\%), and test (20\%) sets.
	
To evaluate the effectiveness of different modeling approaches, we compared the performance of linear regression and Bayesian linear regression models. The results are summarized in \tablename~\ref{table:Table_Linear} and \tablename~\ref{table:Table_Bayes}.

\tablename~\ref{table:Table_Linear} presents the metrics for the linear regression model. The validation mean squared error (MSE) is 171.37, and the test MSE is 156.12. The R-squared values are 0.91 for the validation set and 0.93 for the test set, indicating  91\% of the variance in the validation data and about 93\% in the test data.

\begin{figure}
	\centering
	\includegraphics[width=0.8\columnwidth]{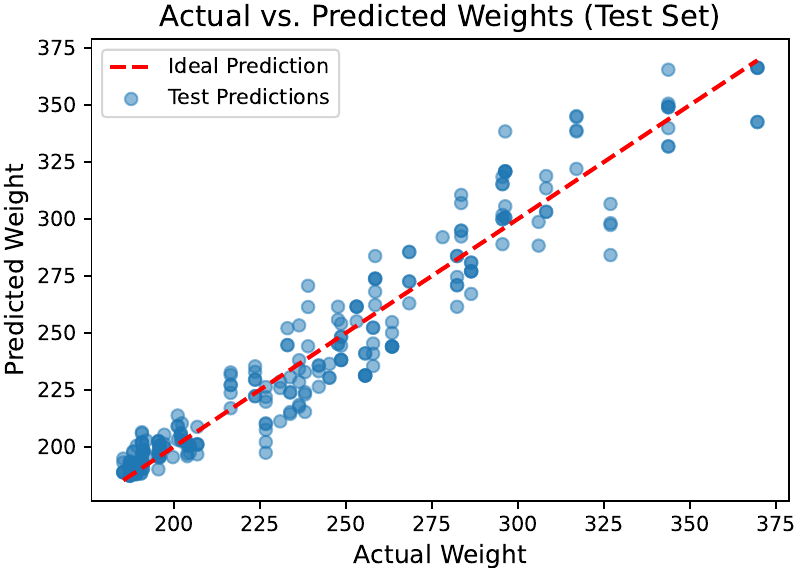}
	\caption{Plot of actual weight vs. predicted weight using linear regression.}
	\label{fig:Plot_linear}
\end{figure}
\begin{figure}
	\centering
	\includegraphics[width=0.8\columnwidth]{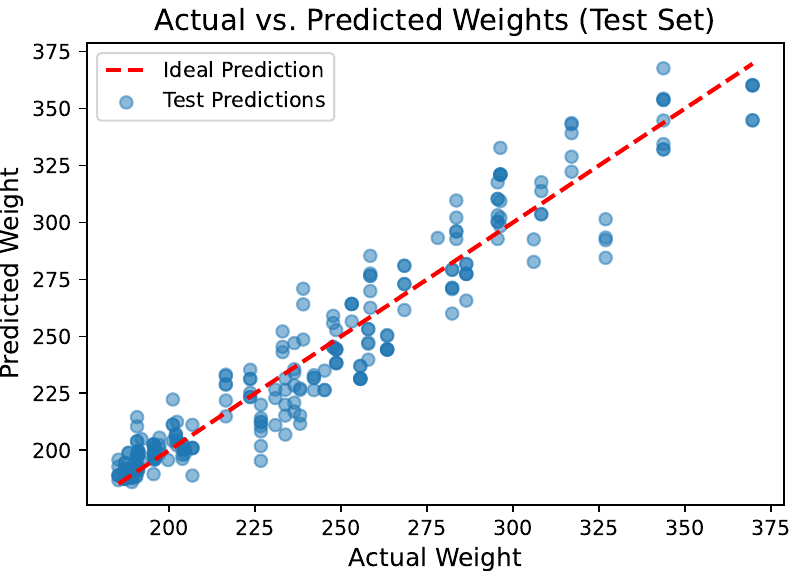}
	\caption{Plot of actual weight vs. predicted weight using Bayes linear regression.}
	\label{fig:Plot_bayes}
\end{figure}

\tablename~\ref{table:Table_Bayes} summarizes the metrics for the Bayesian linear regression model. The validation MSE is 185.61, and the Test MSE is 169.90. The corresponding R-squared values are 0.90 for the validation set and 0.92 for the test set. Although these values are slightly lower than those for the linear regression model, they still explaining over 90\% of the variance in the validation data and about 92\% in the test data.

Both models' predicted versus actual weights are visually represented in \figurename~\ref{fig:Plot_linear} and \figurename~\ref{fig:Plot_bayes}.
In both graphs, the actual weights (x-axis) are plotted against the predicted weights (y-axis) for the test set. The ``Ideal Prediction'' line represents a perfect prediction scenario where the predicted weights would exactly match the actual weights. The ``Test Predictions'' data points show the actual results from the models. These graphs illustrate that both models closely follow the ideal prediction line, indicating their high effectiveness in prediction. 

The comparison between the linear and Bayesian linear regression models shows similar performance metrics in R-squared values, with linear regression performing slightly better in terms of MSE.
This suggests that while the Bayesian model provides a robust framework for handling uncertainty and model complexity, the additional complexity might not translate into significant performance gains in this application.

In summary, both models are highly effective in predicting plant weight, with linear regression showing a slight edge in performance. The high R-squared values across different datasets confirm the models' capability to generalize well across different environmental conditions, making them suitable for practical applications in agricultural settings. Further tests and refinements could focus on enhancing model sensitivity to specific environmental changes and exploring the potential benefits of Bayesian approaches in scenarios with more pronounced data variability and uncertainty.


%
%

\section{Conclusion and Future Work}

In conclusion, this study highlights the value of rapid prototyping in developing advanced CPS for agriculture, promoting sustainable farming practices, and addressing global food demand. Future research should refine CPS technologies and integrate robust data analytics to further boost agricultural productivity and sustainability.
The prototype's adaptability and scalability make it suitable for various agricultural contexts, enhancing sustainable and efficient farming practices. Its real-time environmental data capabilities and precise control over factors like temperature and humidity are crucial.
Additionally, a predictive model using ML algorithms enhances the prototype's ability to forecast plant growth accurately. This blend of rapid prototyping and advanced analytics offers a promising solution for modern agriculture.

For future work, we plan to incorporate more advanced equipment like low-cost depth cameras to replace the LIDAR sensor and camera combination, which could enhance data quality. Simplifying the system's design with a depth camera, which consolidates depth measurement and visual capture, is another potential improvement. Additionally, exploring the effects of different lighting conditions, soil types, hydroponic systems, and nutrient regimes could provide insights into optimal growing conditions.
One major limitation of our ML-based predictive modeling is that the model will be highly dependent on the specific training data.
To overcome this limitation, we plan to incorporate biological models such as what we used~\cite{Niklas1994}.
Developing a fully automated system to adjust environmental parameters in real time could further advance smart farming solutions, creating ideal conditions for plant growth and promoting sustainability in precision agriculture.



\section*{Acknowledgment}
This work was supported in part by the NSF I/UCRC for Intelligent, Distributed, Embedded
Applications and Systems (IDEAS) and from NSF grant \#2231620.


\bibliographystyle{IEEEtran}
\bibliography{library}
\vspace{12pt}
\color{red}

\end{document}